\documentclass[useAMS,usenatbib,usegraphicx,onecolumn]{mn2e}

\usepackage{times}

\def\text{\rm}
\def\Brunt{Brunt-V\"{a}is\"{a}l\"{a}\ }

\title[Probing the internal solar magnetic field through
g-modes]{Probing the internal solar magnetic field through g-modes}

\author[T. I. Rashba, V. B. Semikoz, S. Turck-Chi\`eze and J. W. F. Valle]{%
T. I. Rashba$^{1,2}$\thanks{E-mail: timur@mppmu.mpg.de},
V. B. Semikoz$^{2,3}$\thanks{E-mail: semikoz@ific.uv.es},
S. Turck-Chi\`eze$^{4}$\thanks{E-mail: sylvaine.turck-chieze@cea.fr},
and J. W. F. Valle$^{3}$\thanks{E-mail: valle@ific.uv.es} 
\vspace{0.5cm}
\\
$^{1}$Max-Planck-Institut f\"ur Physik (Werner-Heisenberg-Institut),
F\"ohringer Ring 6, D-80805 Munich, Germany\\
$^{2}$Pushkov Institute of Terrestrial Magnetism, Ionosphere and
Radiowave Propagation of the Russian Academy of Sciences,\\ IZMIRAN,
Troitsk, Moscow region, 142190, Russia\\
$^{3}$AHEP Group, Instituto de F\'isica Corpuscular --
C.S.I.C./Universitat de Val\`encia, Edificio Institutos de Paterna,\\
Apt 22085, E-46071, Val\`encia, Spain,\\
$^{4}$DAPNIA/DSM/Service d'Astrophysique, CEA/Saclay, 91191
Gif-sur-Yvette Cedex, France }

\begin{document}

%\date{\today}
\date{}

\maketitle

%{\noindent  IFIC/06-40, MPP-2006-xxx \hfill}\\

\begin{abstract}
  The observation of g-mode candidates by the SoHO mission opens the
  possibility of probing the internal structure of the solar radiative
  zone (RZ) and the solar core more directly than possible via the use
  of the p-mode helioseismology data. 
We study the effect of rotation and RZ magnetic fields on g-mode
frequencies. 
Using a self-consistent static MHD magnetic field model we show that a
1\% g-mode frequency shift with respect to the Solar Seismic Model
(SSeM) prediction, currently hinted in the GOLF data, can be obtained
for magnetic fields as low as 300 kG, for current measured modes of
radial order $n=-20$.  On the other hand, we also argue that a similar
shift for the case of the low order g-mode candidate ($l=2$, $n=-3$)
frequencies can not result from rotation effects nor from central
magnetic fields, unless these exceed 8 MG.

\end{abstract}


\section{Introduction}

The observation of g-mode candidates by the Global Oscillation at Low
Frequencies (GOLF) instrument aboard the ESA/NASA Solar and
Heliospheric Observatory (SoHO) mission~\citep{TC04} opens new prospects
for solar physics. Indeed, it brings the possibility of probing the
internal structure of the solar radiative zone (RZ) and the solar core
more efficiently and more directly than possible via the use of the
p-mode helioseismology data.

The shift of the g-mode candidate frequencies with respect to the
Solar Seismic Model (SSeM) prediction~\citep{TC01,Couvidat03} hinted
in the experiment is of the order $\delta \omega_{nlm}/\omega_{nl}\sim
1\%$. Other standard models lead to the same kind of discrepancy
but it is clear that none of these models take into account the solar
internal dynamical effects which might contribute to this shift. So
one may attempt to explain this either as resulting from a strong
central magnetic field or by the rotation of the RZ, or both.
Here we study the effect of both RZ magnetic fields and rotation on
g-mode frequencies in order to check previous estimates of their
values~\footnote{{\sl The resulting magnetic and rotation frequency
splittings behave as even (in the Jeffreys-Wentzel-Kramers-Brillouin
(JWKB) approximation used below) and odd with respect to the azimuthal
number $m$, as seen in eqs.~(\ref{shift2}) and (\ref{3Drotation}).}}.

One approach is provided by linearized one-dimensional (1-D)
magneto-hydrodynamics (MHD) \citep{Burgess04a}. This enables us to determine
{\it analytically} the MHD g-mode spectra beyond the JWKB approximation and
opens the possibility of Alfv\'{e}n or slow resonances that could produce
sizeable matter density perturbations in the RZ, and potentially affect solar
neutrino propagation~\citep{Burgess03,Burgess04b}.

Magnetic shifts of g-mode frequencies were calculated using exact
eigenfunctions in 1-D MHD, within the perturbative regime (small
magnetic field)~\citep{Rashba06}. However such approach can not describe
\emph{splittings} in g-mode frequencies, intrinsically associated with
spherical geometry. Moreover, all shifts have the same sign, in
contrast with first indications from experiment.

In this paper we generalize the MHD picture to the three-dimensional
(3-D) case using standard perturbative approach for the calculation of
magnetic field corrections to the g-mode spectra~\citep{Unno, Hasan05}.
The main ingredients of our calculation are: a) a 3-D model of the
background magnetic field in RZ and b) the radial profiles of the
eigenfunctions for the horizontal and radial displacements
$\xi_{h,r}(r)$. For (a) we will use for the first time the static 3-D
background magnetic field solution introduced by Kutvitskii \&
Solov'ev (KS, for short)~\citep{KS}. For (b) we will adopt, as a first
approximation, the eigenfunction calculated for the SSeM neglecting
magnetic fields~\citep{Mathur06} using the Aarhus adiabatic oscillation
package~\citep{Christensen03}.

We organize our presentation as follows. 
In order to calculate the 3-D magnetic shift we first derive in
Section 2 the perturbative magnetic field ${\bf B'}$ using the
explicit 3-D form of the background magnetic field ${\bf B}$ obtained
in a static MHD model~\citep{KS}.  In Section 3 we derive a simple
formula for the magnetic shift in the JWKB approximation (high order
g-modes, $\left| n\right|\gg 1$) and find agreement with calculations
performed for the case of slowly pulsating B-stars~\citep{Hasan05}.
The relevant integrals were calculated numerically using KS background
magnetic field~\citep{KS} and the eigenfunctions $\xi_{r,h}$ were taken
from the SSeM ({\citep{Couvidat03} model 2).
In Section 4 we calculate the rotational splittings of g-modes in
order to compare with those induced by the magnetic field. Finally, in
the Discussion we summarize our results. We conclude that future more
precise g-mode observations will open prospects for narrowing down the
inferred magnetic field strength ranges considerably, when compared
with values currently discussed in the literature, $1~\rmn{G}< B <
30$~MG~\citep{Couvidat03,TC01,Moss03,Ruzmaikin02,Rashba06}.

\section{Magnetic frequency splitting of g modes}

We use the following standard formula for 3-D MHD perturbative
correction (see, e.~g.~\citep{Unno,Hasan05}):
\begin{equation}
  \frac{\delta \omega }{\omega }=\frac{1}{8\pi \omega ^{2}}\frac{\int
    -[(\nabla \times {\bf B^{\prime }})\times {\bf B}+(\nabla \times {\bf B}%
    )\times {\bf B^{\prime }}]\cdot \vec {\xi }^{\ast }dV}{\int \rho \lbrack \xi
    _{r}^{2}+l(l+1)\xi _{h}^{2}]dV},  \label{shift}
\end{equation}
where the eigenvector $\vec{\xi }$ is given by
\begin{equation}\label{xi}
\vec{\xi }=\left[\xi_rY_l^m(\theta, \phi),\xi_h\partial_{\theta}Y_l^m, 
\xi_h\frac{im}{\sin \theta}Y_l^m\right]e^{-i\omega t}.
\end{equation}
The perturbative magnetic field ${\bf B'}$ obtained from Faraday's
equation in ideal MHD, ${\bf B}^{'}= \nabla\times ({\vec{\xi}\times
  {\bf B}})$, takes the form:
\begin{eqnarray}\label{magperturb}
  &&{\bf B'}=\left[({\bf B}\nabla)\xi_r - (\vec{\xi}\nabla)B_r - B_r u\right]{\bf e}_r
  +\nonumber\\
  &&+\left[({\bf B}\nabla)\xi_{\theta} - (\vec{\xi}\nabla)B_{\theta}
  +\frac{(B_{\theta}\xi_r-\xi_{\theta}B_r)}{r}- B_{\theta}u\right]{\bf e}_{\theta}
  +\nonumber\\
  && +\left[({\bf B}\nabla)\xi_{\phi} - (\vec{\xi}\nabla)B_{\phi}
  +\frac{(B_{\phi}\xi_{r}-\xi_{\phi}B_r)}{r}  +
  \frac{(B_{\phi}\xi_{\theta}-\xi_{\phi}B_{\theta})}{r}\cot \theta
  - B_{\phi}u\right]{\bf e}_{\phi},
\end{eqnarray}
where the compressibility $u=\rm {div}~\vec{\xi}$ is given by
$$
u=\frac{\partial \xi_r}{\partial r} + \frac{2}{r}\xi_r +
\frac{1}{r}\frac{\partial \xi_{\theta}}{\partial \theta} + \frac{\cot
  \theta}{r}\xi_{\theta} + \frac{1}{r\sin \theta}\frac{\partial
  \xi_{\phi}}{\partial \phi}~.
$$
Instead of using an \emph{ad hoc} ansatz, we use the static 3-D
configuration for the background magnetic field {\bf B} in the quiet
Sun obtained in~\citep{KS}.  It expresses the equilibrium between the
pressure force, the Lorentz force and the gravitational force,
\begin{equation}\label{equilibrium}
\nabla p -\frac{{\bf j}\times {\bf B}}{c} + \rho \nabla \Phi =0.
\end{equation}
This axisymmetric field
\begin{equation}\label{KSfield}
{\bf B}(r,\theta)=B_{core}\left(\cos \theta b_r^k(x), \sin \theta b_{\theta}^k(x), 
\sin \theta b_{\phi}^k(x) \right)
\end{equation}
is specified as a family of solutions that depend on the roots of the
spherical Bessel functions labeled by $z_k$. These come out from
$f_{5/2}=\sqrt{z}J_{5/2}(z)$ imposing the boundary condition that
${\bf B}$ vanishes on the solar surface, $x=r/R_{\odot}=1$, ${\bf
  B}(R_{\odot})=0$. The amplitude $B_{core}$ characterizes the central
magnetic field strength. The radial and transversal components of
magnetic field are~\citep{KS}:
\begin{eqnarray}\label{profiles}
  &&b_r^k(x)=\frac{1}{(1 - z_k/\sin z_k)}\left[1 - \frac{3}{x^2z_k\sin
      z_k}\left(\frac{\sin (z_kx)}{z_kx} - \cos (z_kx)\right)\right],
  \nonumber\\
  &&  b_{\theta}^k(x)= -\frac{1}{2(1 - z_k/\sin z_k)}\left[2 + \frac{3}{x^2z_k\sin
      z_k}\left(\frac{\sin (z_kx)}{z_kx} - \cos (z_kx) - z_kx\sin
      (z_kx)\right)\right],\nonumber\\
  &&  b_{\phi}^k(x)=\frac{1}{2(1 - z_k/\sin z_k)}\left[z_kx - \frac{3}{x\sin z_k}\left(\frac{\sin (z_kx)}{z_kx} - \cos (z_kx)\right)\right]~.
\end{eqnarray}
Notice that the behavior of {\bf B} at the solar center ($r=0$):
\begin{eqnarray}
&&B_r(0,\theta)=B_{core}\cos \theta,\nonumber\\&&
B_{\theta}(0,\theta)= - B_{core}\sin \theta, \nonumber\\&&
B_{\phi}(0,\theta)=B_{core}\sin \theta\frac{z_k}{2}\frac{r}{R_{\odot}}\to 0,
\end{eqnarray}
is completely regular, determined only by the single parameter
$B_{core}$.

In the left panel of Fig.~\ref{fig:bkutv} we display the perpendicular
component $b_{\perp}^k(r)=\sqrt{(b_{\theta}^k)^2 + (b_{\phi}^k)^2}$
while in the right panel we show the radial $b_r^k(x)$-component given
by Eq.~(\ref{profiles}) for the different modes $k$.
\begin{figure*}
\includegraphics[width=0.7\textwidth]{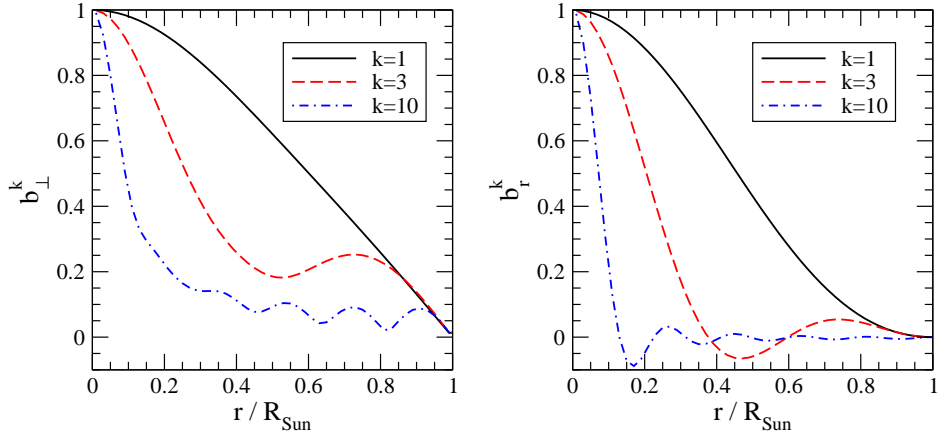}
\caption{\label{fig:bkutv} Perpendicular (left),
  $b_{\perp}^k=\sqrt{(b_\theta^k)^2+(b_\phi^k)^2}$, and radial (right panel),
  $b_r^k$, components of modes $k=1,3,10$ of the self-consistent {\bf
    B}-field~Eq.~(\ref{KSfield}).}
\end{figure*}
Note that such low order modes of the large-scale magnetic
field~Eq.~(\ref{KSfield}), $k=1,2,\dots10$, survive against magnetic
diffusion over the solar age, $\sim 4.6$~Gyr~\citep{Miranda}.

Substituting the background magnetic field in Eq.~(\ref{KSfield}) into
Eq.~(\ref{magperturb}) one gets all components of the perturbative
field ${\bf B}^{'}(r,\theta,\phi)$, as follows:
\begin{eqnarray}\label{magperturb2}
&&B_r^{'}=\frac{B_{core}}{R_{\odot}}\left( \frac{\sin \theta
  \partial_{\theta}Y_l^m}{x}\left[\xi_r(x)b_{\theta}(x) +
  \xi_h(x)b_r(x)\right] +\right.\nonumber\\
&&\left. +Y_l^m\xi_r(x)\left[\frac{imb_{\phi}(x)}{x} - \cos \theta\left(\frac{\partial b_r(x)}{\partial x} + \frac{2b_r(x)}{x}\right)\right] + \frac{l(l+1)\cos \theta Y_l^m}{x}b_r(x)\xi_h(x)\right),
\nonumber\\
&&B_{\theta}^{'}=\frac{B_{core}}{R_{\odot}}\left(\cos \theta
  \partial_{\theta}Y_l^m\left[b_r(x)\frac{\partial \xi_h(x)}{\partial x} -
    \frac{\xi_h(x)}{x}\left(b_{\theta}(x) + b_r(x)\right)\right] +
  \frac{im\partial_{\theta}Y^m_l}{x}\xi_hb_{\phi}-\right.\nonumber\\
&&\left.- \sin \theta Y_l^m\left[
    \frac{b_{\theta}\xi_r}{x} +\frac{\partial (b_{\theta}(x)\xi_r(x))}{\partial
      x}\right] + b_{\theta}\sin
  \theta\left[\frac{\xi_h}{x}\left(\partial^2_{\theta}Y_l^m
      +l(l+1)Y_l^m\right)\right]\right),\nonumber\\
&&B^{'}_{\phi}=\frac{B_{core}}{R_{\odot}}\left(im\cot \theta
  Y_l^m\left[b_r\frac{\partial \xi_h}{\partial x}
    -\frac{\xi_h}{x}\left(2b_{\theta} + b_r\right)\right] +
  im\frac{b_{\theta}\xi_h}{x}\partial_{\theta}Y_l^m -\right.\nonumber\\
&&\left. -\sin \theta Y_l^m\left[\frac{\partial (\xi_r(x)b_{\phi}(x))}{\partial x} + 
  \frac{\xi_rb_{\phi}}{x}\right]+ \frac{\sin \theta
  b_{\phi}\xi_h}{x}\left[l(l+1)Y_l^m +
  \frac{1}{\sin^2\theta}\frac{\partial^2Y}{\partial \phi^2}\right]\right)~.
\end{eqnarray}
Using displacement eigenfunctions $\xi_r^{(nl)}(x)$, $\xi_h^{(nl)}(x)$
for the g-modes of the order $n$ and degree $l$ {\citep{Mathur06}}
we are able to calculate the perturbative magnetic field ${\bf
  B}^{'}(r,\theta,\phi)$ given by Eq.~(\ref{magperturb2}) and then
the integrals in the initial Eq.~(\ref{shift}). This way we obtain
the magnetic shift of eigenfrequencies.

\section{JWKB approximation}

In order to simplify Eq.~(\ref{shift}) we adopt the JWKB well-known
approximation~\citep{Christensen03} (see eqs.~(7.129)-(7.131) there). This holds
for large orders $n\gg 1$ and correspondingly $\omega\ll N$ ($N$ is
the \Brunt frequency). The resulting g-mode eigenfunctions
$\xi_{h}^{(nl)}$, $\xi_{r}^{(nl)}$ of oscillations within RZ are given
by
\begin{equation}\label{xir}
  \xi_r\simeq A\rho^{-1/2}r^{-3/2}\left(\frac{N^2}{\omega^2} -1\right)^{-1/4}\cos \left[\int_{r_1}^r\frac{L}{r^{'}}\left(\frac{N^2}{\omega^2} - 1\right)^{1/2}dr^{'} - \frac{\pi}{4}\right],
\end{equation}
\begin{equation}\label{xih}
  \xi_h\simeq -A\rho^{-1/2}L^{-1}r^{-3/2}\left(\frac{N^2}{\omega^2} -1\right)^{+1/4}\sin \left[\int_{r_1}^r\frac{L}{r^{'}}\left(\frac{N^2}{\omega^2} - 1\right)^{1/2}dr^{'} - \frac{\pi}{4}\right],
\end{equation}
where $L=\sqrt{l(l+1)}$ and the JWKB connection ${\rm d}\xi_r/{\rm
  d}r\simeq L^2\xi_h/r$ was used in obtaining $\xi_h$. 

Obviously, the ratio between the amplitudes of the horizontal and
vertical displacements obeys (see Eq.~(7.132) in~\citep{Christensen03}):
$$
\left|\frac{L\xi_h}{\xi_r}\right|\sim \left(\frac{N^2}{\omega^2} - 1\right)^{1/2}.
$$
showing explicitly that for low degrees $l=1,2$ the g-mode oscillation
is predominantly in the horizontal direction, $\xi_h\gg \xi_r$.
Moreover, differentiating the horizontal component once more,
$\partial \xi_h/\partial r$, one finds that $\partial \xi_h/\partial
r\gg \xi_h/r$, since the additional large factor $[(N/\omega)^2
-1]^{1/2}\gg 1$ appears.

We see that the large derivatives $\partial \xi_h/\partial x$ enter
only in the angular components $B_{\theta}^{'}$ and $B_{\phi}^{'}$
given by Eq.~(\ref{magperturb2}), which with the angular factors
($\cos \theta \partial_{\theta}Y$) and ($\cot \theta Y$) respectively.
As a result, as in Ref.~\citep{Hasan05}, the dominant term in the
integrand of the numerator in Eq.~(\ref{shift}) is the first one
$$
-[(\nabla\times {\bf B}^{'})\times {\bf B}]\cdot \vec{\xi}^{*}=\left| {\bf
    B}^{'}\right|^2\simeq
\left(\frac{B_{core}}{R_{\odot}}\right)^2\left|\frac{1}{x}\frac{{\rm
      d}(xb_r\xi_h)}{{\rm d}x}\right|^2\left[\left| \cos\theta \frac{\partial
      Y}{\partial \theta}\right|^2 + m^2\left|\cot \theta Y\right|^2\right].
$$
The second term in the numerator in Eq.~(\ref{shift}) is negligible
since it is linear in ${\rm d}\xi_h/{\rm d}x$ (in contrast the first
term is quadratic).
Finally, the leading contribution to the denominator is $ \xi_h^2$,
since $\xi_h\gg \xi_r$ for $n\gg 1$.

Under the above approximations we finally obtain the magnetic shift of
g-mode frequencies in the Sun, as
\begin{equation}\label{shift2}
  \frac{\delta \omega}{\omega}=\frac{C_{lm}(B_{core})^2}{8\pi
    \omega^2\rho_cR_{\odot}^2}
  \frac{\int \left| x^{-1}{\rm d}(xb_r\xi_h)/{\rm d}x\right|^2 x^2{\rm d}x}
  {\int\left| \xi_h\right|^2(\rho/\rho_c)x^2{\rm d}x}=S_c\left(\frac{B_{core}}{{\rm MG}}\right)^2.
\end{equation}
Notice that, in contrast to the 1-D case, here there is a dependence
on the angular degree $l$ and azimuthal number $m$. This induces a
splitting (not only a shift) of the g-mode frequencies.
Note also that this expression is in full agreement with the magnetic
shift derived in~\citep{Hasan05} for the g-mode spectra in slowly
pulsating B-stars.  Moreover, the coefficient $C_{lm}$ corresponding to
the angular integration coincides exactly with those of
Ref.~\citep{Hasan05}.
Notice however, that this coincidence was not expected \emph{a priori}
since here we adopt a self-consistent background magnetic
field~\citep{KS} instead of an arbitrary ansatz for $b(x)$, as in
Ref.~\citep{Hasan05}.

In order to compute the magnetic shift we perform the numerical
integration in Eq.~(\ref{shift2}) using the SSeM eigenfunctions
$\xi_{r,h}^{(nl)}$~\citep{Mathur06}  \footnote{Note we use the
  exact eigenfunctions, not the approximations $\xi_{r,h}$ given by
  Eqs.~(\ref{xir}), (\ref{xih}).} shown in Fig.~\ref{fig:gmodes}.
\begin{figure*}
\includegraphics[width=0.7\textwidth]{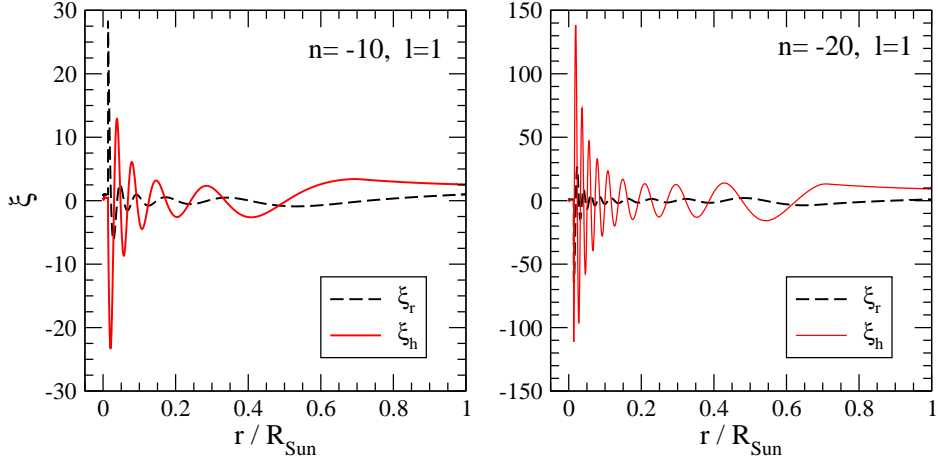}
\caption{\label{fig:gmodes} Horizontal $\xi_h$ and radial $\xi_r$
  eigenfunctions as a function of normalized radius, $r/R_\odot$, are
  shown for modes $g_{10}^1$ (left) and $g_{20}^1$ (right) panel.}
\end{figure*}
Our results are given in Table~1. These results for the magnetic
shifts of the g-mode frequencies in the Sun for $g_{10}^1$ and
$g_{20}^1$.  One finds that magnetic shifts of the magnitude hinted by
the data (order 1\%) can be obtained for RZ magnetic field strength
300 ${\rm kG}$ for the mode $g_{20}^1$ and 2500 ${\rm kG}$ for the
mode $g_{10}^1$. We also give the corresponding values of the $S_c$
coefficient in Eq.~(\ref{shift2}).  Note that the case of young
B-star considered in~\citep{Hasan05} one obtains 110 ${\rm kG}$ and 1100
${\rm kG}$ respectively.
\begin{table}
\begin{center}
  \caption{Solar RZ magnetic field which would produce a frequency of
    1\% for g-modes $g_{20}^1$ and $g_{10}^1$. The magnetic field mode
    $k=1$ is assumed. See text.} % \vskip0.5cm
\begin{tabular}{|l|r|r|r|r|}
\hline
Mode & $\nu$ ($\mu$Hz) & P (hour)  & $S_c$ (MG$^{-2}$) & $B_{core}$ (MG)\\
\hline
$g_{20}^1$ & 33.19 & 8.37 & $1.25\times 10^{-1}$ & 0.28 \\
$g_{10}^1$ & 63.22 & 4.39 & $1.63\times 10^{-3}$ & 2.48\\
\hline
\end{tabular}
\end{center}
\end{table} 

Note that for low order g-mode candidates, e.g.  $g_{3}^2$, ($n=-3$,
$l=2$, $\nu_{3}^2=222.02~\mu$Hz) the JWKB approximation is not
applicable and, moreover, the RZ magnetic field strengths required to
provide the magnetic shift $\sim$ 1\% would be huge, hence outside our
assumptions.

\section{Rotation-induced splitting of g-modes}

Here we give the standard estimate of the rotational splitting of the
g-mode frequencies obtained in the absence of magnetic
field~\citep{Christensen03}. The exact result reads,
\begin{equation}\label{3Drotation}
\delta \omega_{nlm}=m\frac{\int_0^R\Omega (r)[\xi_r^2 + L^2\xi_h^2
-2\xi_r\xi_h -\xi_h^2]r^2\rho{\rm d}r}{\int_0^R[\xi_r^2
+L^2\xi_h^2]r^2\rho{\rm d}r}.
\end{equation}
where the last two terms in the numerator come from the Coriolis force
contribution.

For high order g-modes, $n\gg 1$, for which JWKB approximation is
valid, $\xi_h\gg \xi_r$, and assuming rigid rotation of the RZ,
$\Omega={\rm constant}$, one obtains the simple 3D formula:
\begin{equation}\label{3Dsimple}
\delta \omega_{nlm}=m\beta_{nl}\Omega,
\end{equation}
where the last term in the coefficient $\beta_{nl}=1 - [l(l+1)]^{-1}$
is given by the Coriolis force contribution.

For the lowest degree $l=1$ this gives $\beta_{nl}=1/2$ while for a
high degree $l\gg 1$ the Coriolis term is negligible and
$\beta_{nl}=1$. Choosing $\Omega/2\pi=430$~nHz ($T=27$~d) we see from
the Table~\ref{table2} that the JWKB approximation given by
Eq.~(\ref{3Dsimple}) (fourth column) is in good agreement with the
exact calculations using Eq.~(\ref{3Drotation}) and given in the last
column.

\begin{table}
\begin{center}
\caption{\label{table2}Here we assumed $\Omega=430$\,nHz.}
%\vskip0.5cm
\begin{tabular}{|c|r|r|r|r|}
\hline
     &                 &           & JWKB, 3D              & 3D \\
Mode & $\nu$ ($\mu$Hz) & P (hour)  & $\delta\omega/\omega$ & $\delta\omega/\omega$  \\
&  &  & $ \times m\,10^{-3}$ & $ \times m\, 10^{-3}$  \\
 \hline
 $g_{1}^1$  & 263.07 & 1.05 & $ 0.82 $ &  $0.95$ \\
 $g_{3}^2$  & 222.46 & 1.25 & $ 1.61 $ &  $1.54$ \\
 $g_{4}^1$  & 128.34 & 2.17 & $ 1.68 $ &  $1.64$ \\
 $g_{10}^1$ & 63.22  & 4.39 & $ 3.40$ &  $3.63$ \\
 $g_{10}^2$ & 103.29 & 2.69 & $ 3.47$ &  $3.57$ \\
 $g_{20}^1$ & 33.19  & 8.37 & $ 6.48$ &  $6.60$ \\
 $g_{20}^2$ & 55.98  & 4.96 & $ 6.40$ &  $6.44$ \\
 \hline
 \end{tabular}
 \end{center}
 \end{table} 

\section{Discussion}

Recent observations of g-mode candidates in the GOLF
experiment~\citep{TC04} indicate shifts in the g-mode frequencies
corresponding to low order modes that may be as large as $\sim$ 1\%,
e.~g. $g_1^1$, $g_3^2$.
For such modes the expected rotation splittings seem rather small.  We
have found that $\delta \omega (\mathrm{rotation})/\omega\sim m\times
0.001-0.0015$ for a flat rotation in the radiative zone.  As result
rotation cannot provide the frequency shift hinted by GOLF
observations for these modes. Nevertheless as the rotation in the core
can be slightly increased by a factor greater than 2, it remains
important to properly determine the centroid of the observed modes.
Moreover, it is important to go beyond the standard model picture
and introduce the effect of transport and mixing in the solar
radiative zone which may also impact on the prediction of the high
frequency gravity modes.

Here we mainly study that such shifts may result from RZ magnetic
fields.
In order to give an estimate of the corresponding values of RZ
magnetic fields we have adopted the standard 3-D MHD JWKB perturbative
approach, aware that its validity is limited to relatively large
modes.
We have found that the magnetic shift of g-mode frequencies for high
order modes $\delta \omega/\omega\sim B_{core}^2/\omega^2$ can be of
order 1\% for realistic central magnetic fields values obeying all
known upper bounds discussed in the literature. For example, we have
obtained magnetic fields $B_{core}\simeq 300~{\rm kG}$ and
$B_{core}\simeq 2500~{\rm kG}$ for $g_{20}^1$ and $g_{10}^1$,
respectively. Such values are in qualitative agreement with recent
calculated magnetic field values 110~${\rm kG}$ and 1100 {\rm kG}
providing magnetic shift 1\% for the same modes in the case of young
$4~M_{\odot}$ B-stars~\citep{Hasan05}.
Future observations of such high order g-modes $n\gg 1$ in the Sun
could shed light on the validity of our assumption. In fact we have
already some signature of these modes obtained from their asymptotic
properties. However, due to their very low amplitudes, we have
presently mainly analyzed the sum of dipole gravity modes from $n =
-4$ to $-26$~\citep{Garcia06}. A proper decomposition of the observed
waves will place more constraints on the shift of these modes.

In contrast, to consider the case of low order modes hinted by the
recent GOLF data requires further assumptions. In order to provide
estimates for this case we consider two approximations. First we
recall the general dependence $\delta \omega/\omega\sim
B_{core}^2/\omega^2$. It implies that, in order to provide the same
magnetic shift $\sim $ 1\% the magnetic field for modes $g_1^1$,
$g_3^2$ should be much stronger than in the case of JWKB frequencies,
$n\gg 1$.
E.g. for the g-mode candidate $g_3^2$ with
$\nu=222.46~\mu$Hz~\citep{TC04} from the frequency ratio $\nu
(g_3^2)/\nu (g_{10}^1)=222.46/63.22=3.52$, one finds that the magnetic
field which provides a shift $\sim $ 1\% for $g_3^2$ would be $\sim
8.8$~MG. Such large RZ magnetic field estimate also agrees with the
1-D MHD approach given in Ref.~\citep{Rashba06} (although such 1-D
picture can not provide angular splittings, it has the merit of not
relying on the JWKB approximation).
Indeed, such trend for low order modes, $n=-3,-2,-1$, is seen in Fig.
4 of Ref.~\citep{Rashba06} taking into account that the transversal
wave number parameter ${\rm k=2}$ of the 1-D approach is analogous to
the degree $l$ in the 3-D approach and 2.8 times more magnetic
field value for the maximum \Brunt frequency approximated as
$N=2.8\times 10^{-3}~{\rm rad/sec}=constant$ instead of $\sim
10^{-3}~{\rm rad/sec}$ applied in numerical calculations in~\citep{Rashba06}.

The possible existence of such strong magnetic fields in the RZ is
somewhat disturbing. It could be that neither rotation, nor magnetic
fields are responsible for the hinted frequency shift with respect to
SSeM prediction. 
 For example, the $g_3^2$ mode might mix with other gravity modes
   penetrating RZ with frequencies just below \Brunt frequency.

In this paper we have considered the possibility of a 1\%
   magnetic shift of eigenfrequencies, independently of the frequency
   considered.  The sensitivity to the detailed physics depends
   on the order of the mode as shown in~\citep{Mathur06} and is greater
   at high frequency (low order) than at large order, demonstrating
   the great interest of these modes.  

   The effect of the magnetic field on the frequency shift is expected
   to depend on its shape. In a forthcoming paper we plan to explore
   the sensitivity of magnetic shifts of g-modes in the Sun to the
   structure of magnetic fields given by the self-consistent MHD
   model.  Indeed, refined g-mode data may enable future tomography
   studies of the structure of RZ magnetic fields.

 In short, the search for fundamental solar characteristics deep
 within radiative zone constitutes an important challenge for
 helioseismology methods. Here we have given a conservative limit
   on the magnitude of the magnetic field obtained from its possible
   effect on the g-mode frequencies.  Alternatively, the propagation
 of neutrinos may probe short-wave MHD density perturbations in the
 solar RZ ($\lambda_{MHD}\sim 100-200~{\rm km}$) as already discussed
 in Refs.~\citep{Burgess03,Burgess04a,Burgess04b}. 

\section*{Acknowledgments}
Work supported by MEC grants FPA2005-01269, by ILIAS/N6, EC Contract
Number RII3-CT-2004-506222. TIR was supported by the Marie Curie
Incoming International Fellowship of the European Community. TIR and
VBS acknowledge the Russian Foundation for Basic Research and the RAS
Program ``Solar activity'' for partial support. TIR and VBS thank AHEP
group of IFIC for the hospitality during some stages of the work. VBS
thanks J.-P.  Zahn for fruitful discussions.

 %%%%%%%%%%%%%%%%%%%%%%%%%%%%%%%%%%%%%%%%%%%%


\begin{thebibliography}{99} 
%%%%%%%%%%%%%%%%%%%%%%%%%%%%%%%%%%%%%%%%%%%%%

\bibitem[\protect\citeauthoryear{Burgess et al.}{2003}]{Burgess03}
Burgess, C.P. et al., 2003 , ApJ, 588, L65

\bibitem[\protect\citeauthoryear{Burgess et al.}{2004a}]{Burgess04a}
Burgess, C.P. et al., 2004a, MNRAS, 348, 609

\bibitem[\protect\citeauthoryear{Burgess et al.}{2004b}]{Burgess04b}
Burgess, C.P. et al., 2004b, JCAP, 01, 007

\bibitem[\protect\citeauthoryear{Christensen-Dalsgaard}{2003}]{Christensen03}
Christensen-Dalsgaard, J., 2003, {\em Lecture Notes on Stellar
Oscillations}, available at
{\tt http://astro.phys.au.dk/\~{}jcd/oscilnotes/}

\bibitem[\protect\citeauthoryear{Couvidat, Turck-Chi\`eze \& Kosovichev}{2003}]{Couvidat03} 
Couvidat, S., Turck-Chi\`eze, S., Kosovichev, A.G., 2003, ApJ, 599, 1434

\bibitem[\protect\citeauthoryear{Garcia et al.}{2006}]{Garcia06}
Garc\`ia, R. et al.,
%Turck-Chi\`eze, S., Jim\'enez-Reyes, S. J., Ballot, J., 
%Pall\'e, P. L., Eff-Darwich, A., Mathur, S., Provost, J., 
Science, 2007, in press

\bibitem[\protect\citeauthoryear{Hasan, Zahn \& Christensen-Dalsgaard}{2005}]{Hasan05}
  Hasan, S.S., Zahn J.-P. \& Christensen-Dalsgaard, J. 2005, A\&A 444, L29

\bibitem[\protect\citeauthoryear{Kutvitskii \& Solov'ev}{1994}]{KS}
  Kutvitskii, V.A., Solov'ev L.S., 1994, JETP, 78 (4), 456-464 (In Russian Zh.
  Eksp. Teor. Fiz. 105, 1994, 853-867)

\bibitem[\protect\citeauthoryear{Mathur, Turck-Chi\`eze \& Couvidat}{2006}]{Mathur06}
Mathur, S., Turck-Chi\`eze, S., Couvidat, S. \& Garc\`ia, R.,
2007, ApJ, submitted

\bibitem[\protect\citeauthoryear{Miranda et al.}{2001}]{Miranda}
  Miranda, O.G., et al., 2001, Nucl.Phys., B595, 360-380

\bibitem[\protect\citeauthoryear{Moss}{2003}]{Moss03}
Moss, D. 2003, A\&A, 403, 693

\bibitem[\protect\citeauthoryear{Ruzmaikin \& Lindsey}{2002}]{Ruzmaikin02} 
Ruzmaikin, A., \& Lindsey, C. 2002, in Proceedings of SOHO
12/GONG + 2002 ``Local and Global Helioseismology: The Present and
Future'', Big. Bear Lake, California (USA)

\bibitem[\protect\citeauthoryear{Rashba, Semikoz \& Valle}{2006}]{Rashba06}
 Rashba, T.I., Semikoz, V.B., Valle, J.W.F., 2006, MNRAS, 370, 845-850

\bibitem[\protect\citeauthoryear{Turck-Chi\`eze et al.}{2001}]{TC01}
Turck-Chi\`eze, S. et al., 2001, ApJ, 555, L69

\bibitem[\protect\citeauthoryear{Turck-Chi\`eze et al.}{2004}]{TC04}
Turck-Chi\`eze, S., et al. 2004, ApJ, 604, 455

\bibitem[\protect\citeauthoryear{Unno et al.}{1989}]{Unno}
  Unno, W., Osaki, Y., Ando, H., Saio, H., and Shibahashi, H. 1989, {\it
    Nonradial Oscillations of Stars, 2-nd edition} (Univ. of Tokyo Press,
  Tokyo)

\end{thebibliography}
\end{document}